\documentclass[journal,12pt,onecolumn,draftclsnofoot]{IEEEtran}
\usepackage[latin9]{luainputenc}
\usepackage{float}
\usepackage{amsmath}
\usepackage{graphicx}
\usepackage[unicode=true]
 {hyperref}

\makeatletter

\floatstyle{ruled}
\newfloat{algorithm}{tbp}{loa}
\providecommand{\algorithmname}{Algorithm}
\floatname{algorithm}{\protect\algorithmname}

\usepackage[caption=false,font=footnotesize]{subfig}
\usepackage{algorithm,algpseudocode}
\usepackage{cite}
\usepackage{amssymb}

\usepackage{graphicx}
\usepackage{epsfig}
\usepackage{epstopdf}

\makeatother

\begin{document}

\title{Optimal Portfolio Design for Statistical Arbitrage in Finance}

\author{Ziping~Zhao,~\IEEEmembership{Student Member,~IEEE}, Rui~Zhou~\IEEEmembership{Student Member,~IEEE},
Zhongju~Wang, and Daniel~P.~Palomar,~\IEEEmembership{Fellow,~IEEE}\thanks{This work was supported by the Hong Kong RGC 16208917 research grant.
The work of Z. Zhao was supported by the Hong Kong PhD Fellowship
Scheme (HKPFS).}\thanks{Z. Zhao, R. Zhou, and D. P. Palomar are with the Department of Electronic
and Computer Engineering, The Hong Kong University of Science and
Technology (HKUST), Clear Water Bay, Kowloon, Hong Kong (e-mail: \protect\href{mailto:ziping.zhao@connect.ust.hk}{ziping.zhao@connect.ust.hk};
\protect\href{mailto:rzhouae@connect.ust.hk}{rzhouae@connect.ust.hk};
\protect\href{mailto:palomar@ust.hk}{palomar@ust.hk}).}\thanks{Z. Wang is with the Hong Kong Applied Science and Technology Research
Institute (ASTRI), Hong Kong. (e-mail: \protect\href{mailto:wzhongju@gmail.com}{wzhongju@gmail.com}).}}
\maketitle
\begin{abstract}
In this paper, the optimal mean-reverting portfolio (MRP) design problem
is considered, which plays an important role for the statistical arbitrage
(a.k.a. pairs trading) strategy in financial markets. The target of
the optimal MRP design is to construct a portfolio from the underlying
assets that can exhibit a satisfactory mean reversion property and
a desirable variance property. A general problem formulation is proposed
by considering these two targets and an investment leverage constraint.
To solve this problem, a successive convex approximation method is
used. The performance of the proposed model and algorithms are verified
by numerical simulations.
\end{abstract}


\section{Introduction}

Statistical arbitrage \cite{Pole2011} is a general quantitative investment
and trading strategy widely used by many parties in the financial
markets, e.g., institutional investors, hedge funds, and individual
investors \cite{Krauss2017}. Since it can hedge the overall market
risk, it is also referred to as a market neutral strategy \cite{JacobsLevy2005}.
In statistical arbitrage, the underlying trading basket can consist
of many financial assets of different kinds such as equities, options,
bonds, futures, commodities, etc. In order to arbitrage from the market,
investors should buy the under-priced assets and short-sell the over-priced
ones and profits will be made after the trading positions are unwound
when the ``mis-pricing'' corrects itself. The statistical arbitrage
can be traced back to the famous pairs trading \cite{Vidyamurthy2004}
strategy, a.k.a. spread trading, where only two assets are considered. 

In statistical arbitrage, the trading basket is used to form a ``spread''
characterizing the ``mis-pricing'' of the assets which is stationary,
hence mean-reverting. To make arbitrage, trading is carried out on
the mean reversion (MR) property of the spread, i.e., to buy it when
it is below some statistical equilibrium and sell it when it is above
the statistical equilibrium. There are many ways to design a spread,
like the distance method \cite{GatevGoetzmannRouwenhorst2006}, factor
analysis \cite{AvellanedaLee2010}, and the cointegration method \cite{CaldeiraMoura2013}.
In this paper, we focus on the cointegration method where the spread
is discovered by time series analysis like the ordinary least squares
method in \cite{EngleGranger1987} and the model-based methods in
\cite{Johansen1991,ZhaoPalomar2017a}. In practice, an asset that
naturally shows stationarity is also a spread \cite{ZhangZhang2008}.

The spreads from the statistical estimation methods essentially form
a ``cointegration subspace''. In terms of investment, a natural
question is whether we can design an optimized portfolio from this
subspace. Such a portfolio is named mean-reverting portfolio (MRP).
To design an MRP, there are two objectives to consider: firstly the
MRP should exhibit a strong MR so that it has frequent mean-crossing
points and hence brings in more trading opportunities; and secondly
the designed MRP should exhibit sufficient variance so that each trade
can provide enough profit. These two targets naturally result in a
multi-objective optimization problem, i.e., to find a desirable trade-off
between MR and variance.

In \cite{dAspremont2011}, the author first proposed to design an
MRP by optimizing an MR criterion. Later, authors in \cite{CuturidAspremont2013,CuturidAspremont2016}
found that the method in \cite{dAspremont2011} can result in an MRP
with very low variance, then the variance control was taken into consideration.
But all these works were carried out by using an $\ell_{2}$-norm
constraint on the portfolio weights which do not carry a physical
meaning in finance. To explicitly represent the budget allocation
for different assets, the investment budget constraints were considered
in \cite{ZhaoPalomar2016,ZhaoPalomar2018}. However, in some cases
the methods in \cite{ZhaoPalomar2016,ZhaoPalomar2018} can lead to
very large leverage (i.e., the dollar values employed) which makes
it unacceptable to use for real investment. Besides that, when the
variance is changed, although the investment leverage can change accordingly,
the MR property of the portfolio is insensitive which makes it really
hard to find a desirable trade-off between the MR and the variance
properties in practice.

In this paper, a new optimal MRP design method is proposed that takes
two design objectives and an explicit leverage constraint into consideration.
The objective in this method can suffice to find a desirable trade-off
between the MR and the variance for an MRP. Different MR criteria
are considered and the portfolio constraint takes two cases into consideration.
The design problem finally becomes a nonconvex constrained problem.
A general algorithm based on the successive convex approximation method
(SCA) is proposed. An efficient acceleration scheme is further discussed.
Numerical simulations are carried out to address the efficiency of
the proposed problem model and the solving algorithms. 

\section{Mean-Reverting Portfolio (MRP) Design}

For a financial asset, e.g., a stock, its price at time $t$ is denoted
by $p_{t}$, and its log-price is given by $y_{t}=\log\left(p_{t}\right)$,
where $\log\left(\cdot\right)$ is the natural logarithm. For $M$
assets with log-prices $\mathbf{y}_{t}=\left[y_{1,t},y_{2,t},\ldots,y_{M,t}\right]^{T}$,
one (log-price) spread can be designed by the weights $\boldsymbol{\beta}=\left[\beta_{1},\beta_{2},\ldots,\beta_{M}\right]^{T}$
(say, from the cointegration model) and given by $s_{t}=\boldsymbol{\beta}^{T}\mathbf{y}_{t}$.
Suppose there exists a cointegration subspace with $N$ ($N<M$) cointegration
relations, i.e., $\mathbf{B}=\left[\boldsymbol{\beta}_{1};\boldsymbol{\beta}_{2};\ldots,\boldsymbol{\beta}_{N}\right]$,
then we can have
\begin{equation}
\mathbf{s}_{t}=\mathbf{B}^{T}\mathbf{y}_{t},\label{eq:spread}
\end{equation}
where $\mathbf{s}_{t}$ denote $N$ spreads. Specifically, if the
log-prices are stationary in nature, we get $\mathbf{s}_{t}=\mathbf{y}_{t}$
with $\mathbf{B}=\mathbf{I}$ $\left(N=M\right)$.

The objective of mean-reverting portfolio (MRP) design is to construct
a portfolio of the underlying spreads to attain desirable trading
properties. An MRP is defined by its portfolio weights $\mathbf{w}=\left[w_{1},w_{2},\ldots,w_{N}\right]^{T}$,
with its resulting spread given by $z_{t}=\mathbf{w}^{T}\mathbf{s}_{t}=\sum_{n=1}^{N}w_{n}s_{n,t}$.
Due to \eqref{eq:spread}, we can get
\begin{equation}
\begin{array}{c}
z_{t}=\mathbf{w}_{p}^{T}\mathbf{y}_{t}=\sum_{m=1}^{M}w_{p,m}y_{m,t},\end{array}\label{eq:log-price for MRP}
\end{equation}
where $\mathbf{w}_{p}=\mathbf{B}\mathbf{w}$ are the MRP weights indicating
the market value on different assets. For $m=1,2,\ldots,M$, $w_{p,m}>0$,
$w_{p,m}<0$, and $w_{p,m}=0$ mean a long position (i.e., it is bought),
a short position (i.e., it is short-sold or, more plainly, borrowed
and sold), and no position on the asset, respectively. 

Considering the two design objectives, i.e., MR and variance, we formulate
the optimal MRP design problem as
\begin{equation}
\begin{array}{cl}
\underset{\mathbf{w}}{\text{minimize}} & F\left(\mathbf{w}\right)\triangleq U\left(\mathbf{w}\right)+\mu V\left(\mathbf{w}\right)\\
\text{subject to} & \mathcal{W}=\left\{ \mathbf{w}\mid\|\mathbf{B}\mathbf{w}\|_{1}\leq B\right\} .
\end{array}\label{eq:problem formulation}
\end{equation}
The MR criterion term $U\left(\mathbf{w}\right)$ is jointly represented
as
\[
\begin{array}{l}
\:\:\:U\left(\mathbf{w}\right)\\
=\xi\frac{\mathbf{w}^{T}\mathbf{H}\mathbf{w}}{\mathbf{w}^{T}\mathbf{M}_{0}\mathbf{w}}+\zeta\left(\frac{\mathbf{w}^{T}\mathbf{M}_{1}\mathbf{w}}{\mathbf{w}^{T}\mathbf{M}_{0}\mathbf{w}}\right)^{2}+\eta\sum_{i=2}^{p}\left(\frac{\mathbf{w}^{T}\mathbf{M}_{i}\mathbf{w}}{\mathbf{w}^{T}\mathbf{M}_{0}\mathbf{w}}\right)^{2},
\end{array}
\]
which particularizes to the predictability statistics $\mathrm{pre}\left(\mathbf{w}\right)$
with $\xi=1$, $\mathbf{H}=\mathbf{M}_{1}^{T}\mathbf{M}_{0}^{-1}\mathbf{M}_{1}$,
and $\zeta=\eta=0$; the portmanteau statistics $\mathrm{por}\left(p,\mathbf{w}\right)$
with $\xi=0$, and $\zeta=\eta=1$; the crossing statistics $\mathrm{pre}\left(\mathbf{w}\right)$
with $\xi=1$, $\mathbf{H}=\mathbf{M}_{1}$, and $\zeta=\eta=0$;
and the penalized crossing statistics $\mathrm{pcro}\left(p,\mathbf{w}\right)$
with $\xi=1$, $\mathbf{H}=\mathbf{M}_{1}$, $\zeta=0$, and $\eta>0$,
where $\mathbf{M}_{i}=\mathsf{Cov}\left(\mathbf{s}_{t},\mathbf{s}_{t+i}\right)$
for $i=1,\ldots,p$ \cite{CuturidAspremont2013,ZhaoPalomar2018}.
The variance term $V\left(\mathbf{w}\right)$ is represented by
\[
\begin{array}{c}
V\left(\mathbf{w}\right)=\frac{1}{\mathbf{w}^{T}\mathbf{M}_{0}\mathbf{w}}.\end{array}
\]
And $\mu\geq0$ defines the trade-off between the MR and variance.
Specially, when $\mu=0$, the designed MRP has the best MR property;
and likewise when $\mu\rightarrow\infty$, the problem leads to the
MRP with best variance. In the constraint set $\mathcal{W}$, $B$
means the total leverage deployed on all the assets in an investment.
The problem in \eqref{eq:problem formulation} is a nonconvex constrained
problem with a nonconvex smooth objective and a convex nonsmooth constraint.

\section{Problem Solving via The SCA Method}

\subsection{The Successive Convex Approximation Method}

The successive convex approximation (SCA) method \cite{MarksWright1978}
is a general optimization method especially for nonconvex problems.
In this paper, a variant of SCA in \cite{ScutariFacchineiSongPalomarPang2014}
is used, which solves the original problem by solving a sequence of
strongly convex problems and can also preserve feasibility of the
iterates. Specifically, a problem is given as follows:
\begin{equation}
\begin{array}{cccc}
\underset{\mathbf{x}}{\mathsf{minimize}} & f\left(\mathbf{x}\right) & \mathsf{subject\:to} & \mathbf{x}\in{\cal X},\end{array}\label{eq:SCA}
\end{equation}
where ${\cal X}\subseteq\mathbb{R}^{N}$ and no assumption is on the
convexity and smoothness of $f\left(\mathbf{x}\right)$ and ${\cal X}$.
Instead of tackling \eqref{eq:SCA} directly, starting from an initial
point $\mathbf{x}^{\left(0\right)}$, the SCA method solves a series
of subproblems with surrogate functions $\tilde{f}\left(\mathbf{x};\mathbf{x}^{\left(k\right)}\right)$
approximating the original objective $f\left(\mathbf{x}\right)$ and
a sequence $\left\{ \mathbf{x}^{\left(k\right)}\right\} $ is generated
by the following rules:
\begin{equation}
\begin{cases}
\begin{array}{l}
\hat{\mathbf{x}}^{\left(k+1\right)}=\arg\underset{\mathbf{x}\in\mathcal{X}}{\min}\tilde{f}\left(\mathbf{x};\mathbf{x}^{\left(k\right)}\right)\\
\mathbf{x}^{\left(k+1\right)}=\mathbf{x}^{\left(k\right)}+\gamma^{\left(k\right)}\left(\hat{\mathbf{x}}^{\left(k+1\right)}-\mathbf{x}^{\left(k\right)}\right).
\end{array}\end{cases}\label{eq:SCA update}
\end{equation}
The first step is to generate the descent direction (i.e., $\hat{\mathbf{x}}^{\left(k+1\right)}-\mathbf{x}^{\left(k\right)}$)
by solving a best-response problem, and the second step is the variable
update with step-size $\gamma^{\left(k\right)}$. For $\tilde{f}\left(\mathbf{x};\mathbf{x}^{\left(k\right)}\right)$,
the following conditions are needed:

\textbf{A1)} given $\mathbf{x}^{\left(k\right)}$, $\tilde{f}\left(\mathbf{x};\mathbf{x}^{\left(k\right)}\right)$
is $c$-strongly convex on $\mathcal{X}$ for some $c>0$, i.e., $\nabla_{\mathbf{x}}^{2}\tilde{f}\left(\mathbf{x};\mathbf{x}^{\left(k\right)}\right)\succeq c\mathbf{I}$;

\textbf{A2)} $\nabla_{\mathbf{x}}\tilde{f}\left(\mathbf{x}^{\left(k\right)};\mathbf{x}^{\left(k\right)}\right)=\nabla_{\mathbf{x}}f\left(\mathbf{x}^{\left(k\right)}\right)$
for all $\mathbf{x}^{\left(k\right)}\in\mathcal{X}$

\textbf{A3)} $\nabla_{\mathbf{x}}\tilde{f}\left(\mathbf{x};\mathbf{x}\right)$
is continuous for all $\mathbf{x}\in\mathcal{X}$.

It is easy to see that the key point in SCA is to find a good approximation
$\tilde{f}\left(\mathbf{x};\mathbf{x}^{\left(k\right)}\right)$ and
to choose a proper step-size $\gamma^{\left(k\right)}$ for a fast
convergence.

\subsection{Optimal MRP Design Based on The SCA Method}

Applying the SCA method to solve problem \eqref{eq:problem formulation},
we can first have the convex approximation function $\tilde{F}(\mathbf{w};\mathbf{w}^{\left(k\right)})$
given by
\begin{equation}
\begin{array}{l}
\;\;\tilde{F}\left(\mathbf{w};\mathbf{w}^{\left(k\right)}\right)\\
=\tilde{U}\left(\mathbf{w};\mathbf{w}^{\left(k\right)}\right)+\mu\tilde{V}\left(\mathbf{w};\mathbf{w}^{\left(k\right)}\right)+\tau\|\mathbf{w}-\mathbf{w}^{\left(k\right)}\|_{2}^{2},
\end{array}\label{eq:partial lin}
\end{equation}
with the parameter $\tau\geq0$ on the proximal term is added for
convergence reason. To get the convex approximation $\tilde{U}\left(\mathbf{w};\mathbf{w}^{\left(k\right)}\right)$,
the second and the third nonconvex terms in $U\left(\mathbf{w}\right)$
are convexified by linearizing each term inside the squares $\left(\cdot\right)^{2}$.
This approximation technique ensures the same gradient for $\tilde{U}\left(\mathbf{w};\mathbf{w}^{\left(k\right)}\right)$
with $U\left(\mathbf{w}\right)$ and naturally keep the convex structure
$U\left(\mathbf{w}\right)$. Then $\tilde{U}\left(\mathbf{w};\mathbf{w}^{\left(k\right)}\right)$
is given by
\begin{equation}
\begin{array}{c}
\tilde{U}\left(\mathbf{w};\mathbf{w}^{\left(k\right)}\right)=\mathbf{w}^{T}\mathbf{A}_{\mathrm{U}}^{\left(k\right)}\mathbf{w}+\mathbf{b}_{\mathrm{U}}^{\left(k\right)T}\mathbf{w},\end{array}\label{eq:partial lin for U}
\end{equation}
where $\mathbf{A}_{\mathrm{U}}^{\left(k\right)}\triangleq4\zeta(\mathbf{d}_{0,1}^{\left(k\right)}\mathbf{d}_{0,1}^{\left(k\right)T}+\mathbf{d}_{1,0}^{\left(k\right)}\mathbf{d}_{1,0}^{\left(k\right)T}-\mathbf{d}_{0,1}^{\left(k\right)}\mathbf{d}_{1,0}^{\left(k\right)T}-\mathbf{d}_{1,0}^{\left(k\right)}\mathbf{d}_{0,1}^{\left(k\right)T})+4\eta\sum_{i=2}^{p}(\mathbf{d}_{0,i}^{\left(k\right)}\mathbf{d}_{0,i}^{\left(k\right)T}+\mathbf{d}_{i,0}^{\left(k\right)}\mathbf{d}_{i,0}^{\left(k\right)T}-\mathbf{d}_{0,i}^{\left(k\right)}\mathbf{d}_{i,0}^{\left(k\right)T}-\mathbf{d}_{i,0}^{\left(k\right)}\mathbf{d}_{0,i}^{\left(k\right)T})$,
and $\mathbf{b}_{\mathrm{U}}^{\left(k\right)}\triangleq2\xi(\mathbf{d}_{0,h}^{\left(k\right)}-\mathbf{d}_{h,0}^{\left(k\right)})+2\zeta r_{1}^{\left(k\right)}(\mathbf{d}_{0,1}^{\left(k\right)}-\mathbf{d}_{1,0}^{\left(k\right)})+2\eta\sum_{i=2}^{p}r_{i}^{\left(k\right)}(\mathbf{d}_{0,i}^{\left(k\right)}-\mathbf{d}_{i,0}^{\left(k\right)})$,
with $r_{h}^{\left(k\right)}=(\mathbf{w}^{\left(k\right)T}\mathbf{H}\mathbf{w}^{\left(k\right)})$
$/(\mathbf{w}^{\left(k\right)T}\mathbf{M}_{0}\mathbf{w}^{\left(k\right)})$,
$r_{i}^{\left(k\right)}=(\mathbf{w}^{\left(k\right)T}\mathbf{M}_{i}\mathbf{w}^{\left(k\right)})/(\mathbf{w}^{\left(k\right)T}\mathbf{M}_{0}\mathbf{w}^{\left(k\right)})$,
$\mathbf{d}_{0,h}^{\left(k\right)}=\mathbf{H}\mathbf{w}^{\left(k\right)}/(\mathbf{w}^{\left(k\right)T}\mathbf{M}_{0}\mathbf{w}^{\left(k\right)})$,
$\mathbf{d}_{h,0}^{\left(k\right)}=r_{h}^{\left(k\right)}\mathbf{M}_{0}\mathbf{w}^{\left(k\right)}$,
$\mathbf{d}_{0,i}^{\left(k\right)}=\mathbf{M}_{i}\mathbf{w}^{\left(k\right)}$
$/(\mathbf{w}^{\left(k\right)T}\mathbf{M}_{0}\mathbf{w}^{\left(k\right)})$,
and $\mathbf{d}_{i,0}^{\left(k\right)}=r_{i}^{\left(k\right)}\mathbf{M}_{0}\mathbf{w}^{\left(k\right)}$,
with $i=1,\ldots,p$. Likewise, the $\tilde{V}\left(\mathbf{w};\mathbf{w}^{\left(k\right)}\right)$
is the convex approximation for the variance term $V\left(\mathbf{w}\right)$
which is given by
\begin{equation}
\begin{array}{c}
\tilde{V}\left(\mathbf{w};\mathbf{w}^{\left(k\right)}\right)=\mathbf{b}_{\mathrm{V}}^{\left(k\right)T}\mathbf{w},\end{array}\label{eq:partial lin for V}
\end{equation}
where $\mathbf{b}_{\mathrm{V}}^{\left(k\right)}\triangleq-2\left(\mathbf{w}^{\left(k\right)T}\mathbf{M}_{0}\mathbf{w}^{\left(k\right)}\right)^{-2}\mathbf{M}_{0}\mathbf{w}^{\left(k\right)}$. 

Then by combining $\tilde{U}\left(\mathbf{w};\mathbf{w}^{\left(k\right)}\right)$
with $\tilde{V}\left(\mathbf{w};\mathbf{w}^{\left(k\right)}\right)$
and dropping some constants, the $\tilde{F}\left(\mathbf{w};\mathbf{w}^{\left(k\right)}\right)$
in \eqref{eq:partial lin} becomes
\begin{equation}
\tilde{F}\left(\mathbf{w};\mathbf{w}^{\left(k\right)}\right)=\mathbf{w}^{T}\mathbf{A}^{\left(k\right)}\mathbf{w}+\mathbf{b}^{\left(k\right)T}\mathbf{w},\label{eq:partial lin A b}
\end{equation}
where $\mathbf{A}^{\left(k\right)}\triangleq\mathbf{A}_{\mathrm{U}}^{\left(k\right)}+\tau\mathbf{I}$,
and $\mathbf{b}^{\left(k\right)}\triangleq\mathbf{b}_{\mathrm{U}}^{\left(k\right)}+\mu\mathbf{b}_{\mathrm{V}}^{\left(k\right)}-2\tau\mathbf{w}^{\left(k\right)}$.
And the subproblem to solve becomes

\begin{equation}
\begin{array}{cl}
\underset{\mathbf{w}}{\text{minimize}} & \mathbf{w}^{T}\mathbf{A}^{\left(k\right)}\mathbf{w}+\mathbf{b}^{\left(k\right)T}\mathbf{w}\\
\text{subject to} & \|\mathbf{B}\mathbf{w}\|_{1}\leq B,
\end{array}\label{eq:subproblem in SCA}
\end{equation}
which is a convex problem. Summarizing, in order to solve the original
problem \eqref{eq:problem formulation}, we just need to iteratively
solve a sequence of convex problems \eqref{eq:subproblem in SCA}.
The SCA-based algorithm is named SCA-MRP and given in Algorithm \ref{alg:SCA_MRP}.
The convergence of Algorithm \ref{alg:SCA_MRP} can be guaranteed
if the step-size $\gamma^{\left(k\right)}$ is chosen as a (suitably
small) constant or alternatively chosen according to the following
diminishing step-size rule:
\[
\begin{array}{cl}
\text{Given} & \gamma^{\left(0\right)}\in\left(0,1\right],\\
\text{Let} & \gamma^{\left(k+1\right)}=\gamma^{\left(k\right)}\left(1-\theta\gamma^{\left(k\right)}\right)\;\text{for }k=0,1,2,\ldots,
\end{array}
\]
where $\theta\in\left(0,1\right)$ is a given constant. 

\begin{algorithm}[H]
\begin{algorithmic}[1]
\Require $\mathbf{H}$, $\mathbf{M}_{i}$ $(i=0,\ldots,p)$, $\mu$, $\mathbf{B}$, $B$ and $\tau$
\State Set   $k=0$, $\gamma^{(0)}$ and $\mathbf{w}^{(0)}$.

\Repeat

\State Compute $\mathbf{A}^{(k)}$ and $\mathbf{b}^{(k)}$.

\State $\hat{\mathbf{w}}^{(k+1)} = \arg\underset{\mathbf{w}\in\mathcal{W}}{\min}\mathbf{w}^{T}\mathbf{A}^{(k)}\mathbf{w}+\mathbf{b}^{(k)T}\mathbf{w}$
		
\State $\mathbf{w}^{(k+1)}=\mathbf{w}^{(k)}+\gamma^{(k)}(\hat{\mathbf{w}}^{(k+1)} - \mathbf{w}^{(k)})$
		
\State $k\gets k+1$

\Until convergence

\end{algorithmic}\caption{\label{alg:SCA_MRP}SCA-MRP Algorithm for Optimal MRP Design}
\end{algorithm}

The inner convex problem (see $\mathtt{Step}$ $\mathtt{5}$ in Algorithm
\ref{alg:SCA_MRP}) has no closed-form solution, but we can resort
to the off-the-shelf solvers like $\mathtt{MOSEK}$ \cite{Mosek2015}
or the popular scripting language $\mathtt{CVX}$ \cite{GrantBoydYe2008}.
However, as an alternative to the general-purpose methods, we can
also develop problem-specific algorithms to solve this problem efficiently.

\subsection{Solving The Inner Subproblem in SCA Using ADMM}

The alternating direction method of multipliers (ADMM) is widely used
to solve convex problems by breaking them into smaller parts, each
of which is then easier to handle \cite{BoydParikhChuPeleatoEckstein2011}.
To solve problem \eqref{eq:subproblem in SCA} by ADMM, we first rewrite
problem \eqref{eq:subproblem in SCA} (for notational simplicity,
superscripts $\left(k\right)$ are omitted) by introducing an auxiliary
variable $\mathbf{z}=\mathbf{B}\mathbf{w}$ as follows:
\begin{equation}
\begin{array}{cl}
\underset{\mathbf{w},\mathbf{z}}{\text{minimize}} & \mathbf{w}^{T}\mathbf{A}\mathbf{w}+\mathbf{b}^{T}\mathbf{w}\\
\text{subject to} & \|\mathbf{z}\|_{1}\leq B,\;\mathbf{B}\mathbf{w}-\mathbf{z}=\mathbf{0}.
\end{array}\label{eq:subproblem-ADMM}
\end{equation}
Then the augmented Lagrangian is
\[
\begin{array}{l}
\;\;L_{\rho}\left(\mathbf{w},\mathbf{z},\mathbf{u}\left(\mathbf{y}\right)\right)\\
=\mathbf{w}^{T}\mathbf{A}\mathbf{w}+\mathbf{b}^{T}\mathbf{w}+I_{\mathcal{C}}(\mathbf{z})+\frac{\rho}{2}\left\Vert \mathbf{P}\mathbf{w}-\mathbf{z}+\mathbf{u}\right\Vert _{2}^{2},
\end{array}
\]
where $I_{\mathcal{C}}\left(\mathbf{z}\right)=\left\{ \begin{array}{l}
0,\quad\mathbf{z}\in\mathcal{C}\\
+\infty,\text{otherwise}
\end{array}\right.$ with $\mathcal{C}\triangleq\left\{ \mathbf{z}\big|\left\Vert \mathbf{z}\right\Vert _{1}\leq B\right\} $
is the indicator function and the penalty parameter $\rho>0$ serves
as the dual update step-size with the scaled dual variable $\mathbf{u}=\frac{1}{\rho}\mathbf{y}$.
Then, the ADMM updates are given in three variable blocks, i.e., ($\mathbf{w},\mathbf{z},\mathbf{u}\left(\mathbf{y}\right)$),
by
\[
\begin{cases}
\begin{array}{l}
\mathbf{w}^{\left(k+1\right)}=\arg\underset{\mathbf{w}}{\min}\Bigl\{\mathbf{w}^{T}\mathbf{A}\mathbf{w}+\mathbf{b}^{T}\mathbf{w}\\
\quad\quad\quad\quad\quad\quad\quad\quad\quad\quad\quad\quad+\frac{\rho}{2}\left\Vert \mathbf{B}\mathbf{w}-\mathbf{z}^{\left(k\right)}+\mathbf{u}^{\left(k\right)}\right\Vert _{2}^{2}\Bigr\}\\
\mathbf{z}^{\left(k+1\right)}=\arg\underset{\mathbf{z}}{\min}\left\{ I_{\mathcal{C}}(\mathbf{z})+\frac{\rho}{2}\left\Vert \mathbf{z}-\mathbf{B}\mathbf{w}^{\left(k+1\right)}-\mathbf{u}^{\left(k\right)}\right\Vert _{2}^{2}\right\} \\
\mathbf{u}^{\left(k+1\right)}=\mathbf{u}^{\left(k\right)}+\mathbf{B}\mathbf{w}^{\left(k+1\right)}-\mathbf{z}^{\left(k+1\right)}.
\end{array}\end{cases}
\]
Specifically, for variable $\mathbf{w}$, it is to solve a convex
quadratic programming with a closed-form solution as follows:
\[
\mathbf{w}^{\left(k+1\right)}=-\left(2\mathbf{A}+\rho\mathbf{B}^{T}\mathbf{B}\right)^{-1}\left(\mathbf{b}+\rho\mathbf{B}^{T}\left(\mathbf{u}^{\left(k\right)}-\mathbf{z}^{\left(k\right)}\right)\right).
\]
By defining $\mathbf{h}^{\left(k\right)}=\mathbf{B}\mathbf{w}^{\left(k+1\right)}+\mathbf{u}^{\left(k\right)}$,
the variable $\mathbf{z}$ update is equivalent to solve
\begin{equation}
\begin{array}{c}
\mathbf{z}^{\left(k+1\right)}=\arg\underset{\mathbf{z}\in\mathcal{C}}{\min}\left\Vert \mathbf{z}-\mathbf{h}^{\left(k\right)}\right\Vert _{2}^{2}=\Pi_{\mathcal{C}}\left(\mathbf{h}^{\left(k\right)}\right),\end{array}\label{eq:Proj_L1_Ball}
\end{equation}
which is the classical projection onto the $\ell_{1}$-ball problem
\cite{Palomar2005,DuchiShalev-ShwartzSingerChandra2008} with $\Pi_{\mathcal{C}}\left(\cdot\right)$
denoting the projection operator. This problem has a closed-form solution
given in the following.

\begin{algorithm}
\begin{algorithmic}
\If {$||\mathbf{h}||_1 \leq B$}
\State $\mathbf{z} = \mathbf{h}$, \Return $\mathbf{z}$
\Else
\State $\mathbf{a} = \mathrm{sign}(\mathbf{h})$ and $\mathbf{b} = \mathrm{abs}(\mathbf{h})$ 
\State Sort $\mathbf{b}$ in order: $b_{(1)} \geq b_{(2)} \geq \dots \geq b_{(N)}$ 
\State $\rho=\arg\underset{1 \leq j \leq N}{\max} \bigg\{ b_{(j)} - \frac{1}{j} \Big(\sum_{i=1}^{j} b_{(i)} - B\Big) >0 \bigg\} $ 
\State $\theta = \frac{1}{\rho} \Big( \sum_{i=1}^{\rho} b_{(i)} - B \Big)$ 
\State $z_{j} = a_{j} \max \{b_{j} - \theta, 0\}$, $1 \leq j \leq N$, \Return $\mathbf{z}$ 
\EndIf
\end{algorithmic}

\caption{Projection onto the $\ell_{1}$-ball.}
\end{algorithm}

In the solution for $\mathbf{z}$-update, $\mathrm{sgn}\left(\cdot\right)$
is the ``sign function''; $\mathrm{abs}\left(\cdot\right)$ is the
absolute value function; and $b_{(j)}$ $\left(1\leq j\leq N\right)$
denotes the $j$-th largest element in $\mathbf{b}$. Then, the overall
ADMM-based algorithm can be summarized in Algorithm \ref{alg:ADMM_MRP_SUBPROBLEM}.

\begin{algorithm}[t]
\begin{algorithmic}[1]
\Require $\mathbf{A}$, $\mathbf{b}$, $\mathbf{B}$,  $B$ and $\rho$

\State Set $\mathbf{w}^{(0)}$, $\mathbf{z}^{(0)}$, $\mathbf{u}^{(0)}$ and $k=0$.
\Repeat		
\State $\mathbf{w}^{(k+1)} = -(2\mathbf{A}+\rho\mathbf{B}^{T}\mathbf{B})^{-1}(\mathbf{b}+\rho\mathbf{B}^{T}(\mathbf{u}^{(k)}-\mathbf{z}^{(k)}))$
\State $\mathbf{h}^{(k)}=\mathbf{B}\mathbf{w}^{(k+1)}+\mathbf{u}^{(k)}$
\State $\mathbf{z}^{(k+1)} = \Pi_{\mathcal{C}}(\mathbf{h}^{(k)})$ 		
\State $\mathbf{u}^{(k+1)} = \mathbf{u}^{(k)}+\mathbf{B}\mathbf{w}^{(k+1)}-\mathbf{z}^{(k+1)}$ 		
\State $k\gets k+1$
\Until convergence	

\end{algorithmic}\caption{\label{alg:ADMM_MRP_SUBPROBLEM}An ADMM-Based Algorithm for Problem
\eqref{eq:subproblem in SCA}}
\end{algorithm}

\subsection{Acceleration Scheme for The SCA-MRP Algorithm \label{sec:Acceleration-for-SCA}}

Besides the diminishing step-size rules for step-size $\gamma^{\left(k\right)}$,
it is possible to get a better convergence speed for Algorithm \ref{alg:SCA_MRP}
by using some acceleration method. One Armijo-like backtracking line
search\textbf{ }rule for the step-size \cite{ScutariFacchineiLampariello2017}
is given as follows:
\[
\begin{array}{cl}
\mathrm{Given} & \alpha,\beta\in\left(0,1\right),\;l=0\\
\mathrm{While} & \Delta F\left(\mathbf{w}^{\left(k\right)}\right)>-\alpha\beta^{l}\|\Delta\mathbf{w}^{\left(k\right)}\|_{2}^{2}\\
 & \quad l=l+1\\
\mathrm{Let} & \gamma^{\left(k\right)}=\beta^{l}\;\text{for }k=0,1,2,\ldots,
\end{array}
\]
where $\Delta F\left(\mathbf{w}^{\left(k\right)}\right)=F\left(\mathbf{w}^{\left(k\right)}+\beta^{l}\Delta\mathbf{w}^{\left(k\right)}\right)-F\left(\mathbf{w}^{\left(k\right)}\right)$
with $\Delta\mathbf{w}^{\left(k\right)}=\hat{\mathbf{w}}^{\left(k+1\right)}-\mathbf{w}^{\left(k\right)}$.

\section{Numerical Simulations\label{sec:Numerical-Simulations}}

\begin{figure}[H]
\begin{centering}
\includegraphics[width=0.5\columnwidth]{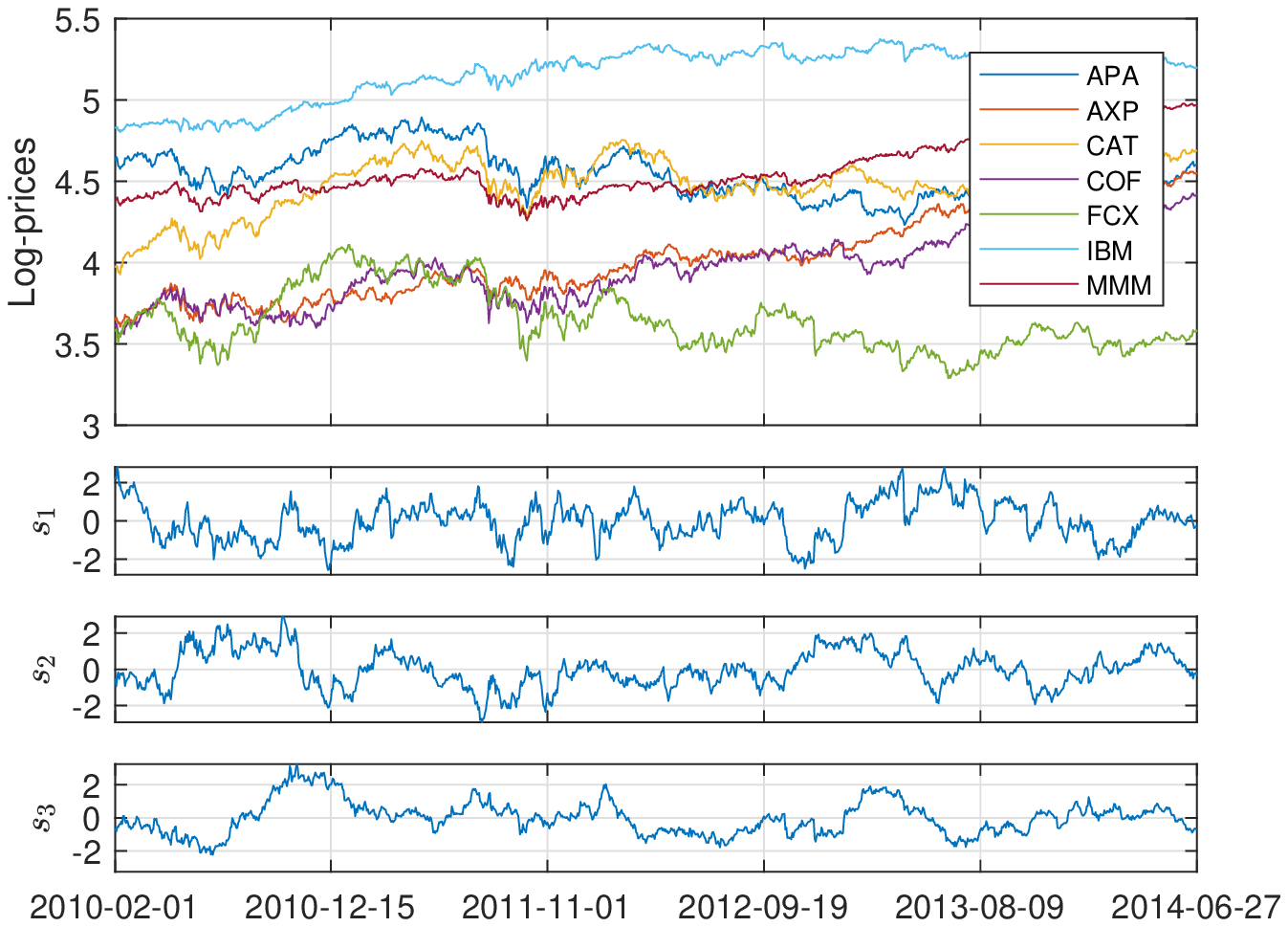}\includegraphics[width=0.5\columnwidth]{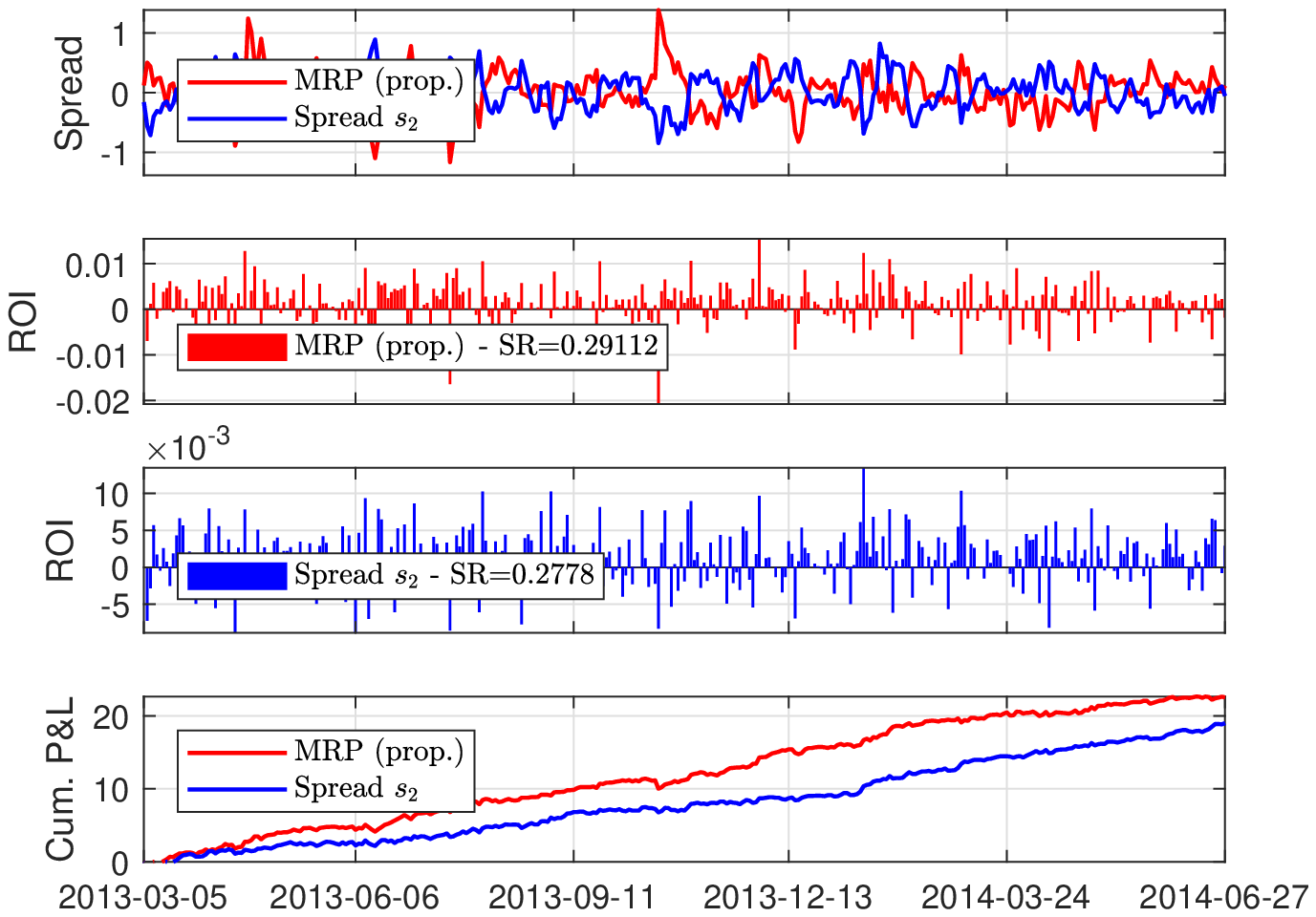}
\par\end{centering}
\caption{\label{fig:trading}A mean-reversion trading based on real data.}
\end{figure}

In this section, we test the proposed problem formulation and algorithms
using market data from the Standard \& Poor's 500 (S\&P 500) Index,
which are retrieved from Google Finance\footnote{\href{https://www.google.com/finance}{https://www.google.com/finance}}.
We choose stock candidates into one asset pool as \{$\mathsf{APA}$,
$\mathsf{AXP}$, $\mathsf{CAT}$, $\mathsf{COF}$, $\mathsf{FCX}$,
$\mathsf{IBM}$, $\mathsf{MMM}$\}, where they are denoted by their
ticker symbols in Figure \ref{fig:trading}. Three spreads are constructed
from this pool based on the Johansen method as shown in Figure \ref{fig:trading}.
The methods proposed in this paper are employed for optimal MRP design.
After that, we apply the designed MRP to a mean reversion trading
based on the trading framework and performance measure introduced
in \cite{ZhaoPalomar2018}. In Figure \ref{fig:trading}, we compare
the performance of our designed MRP with spread $s_{2}$. The performance
metrics like return on investment (ROI), Sharpe ratio, and cumulative
P\&Ls are reported. It is shown that the designed MRP can achieve
a higher Sharpe ratio and a better final cumulative return.

We further show the convergence property over iterations of the objective
function value in problem \eqref{eq:problem formulation} by using
the proposed SCA-MRP algorithm with and without acceleration in comparison
to some benchmark algorithms.

\begin{figure}[H]
\begin{centering}
\includegraphics[width=0.6\columnwidth]{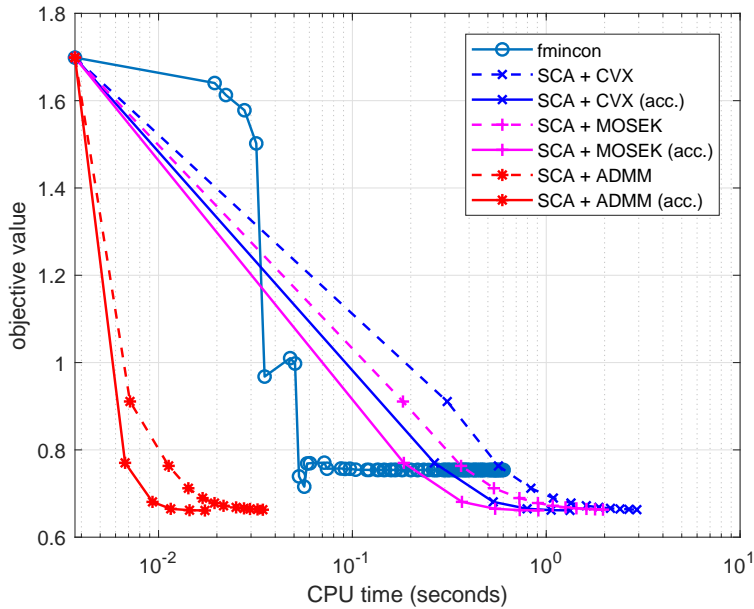}
\par\end{centering}
\centering{}\caption{\label{fig:Convergence-comparison}Convergence comparison for objective
function value.}
\end{figure}

\begin{figure}[H]
\centering{}\includegraphics[width=0.6\columnwidth]{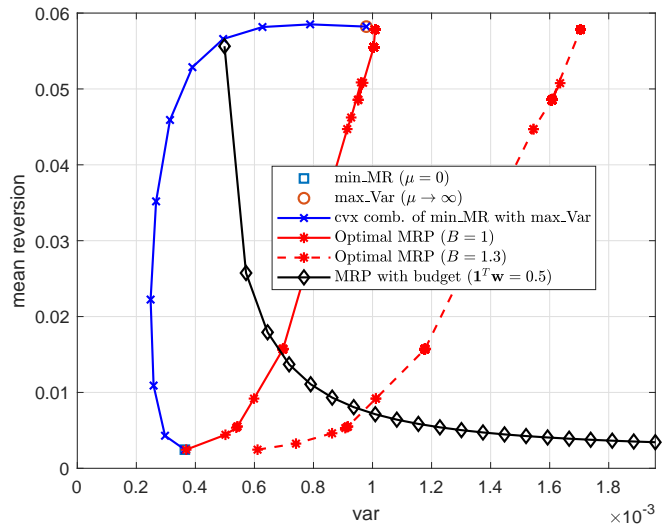}\caption{\label{fig:MR-Var}Trade-off between MR and variance in MRP design.}
\end{figure}

The SCA-MRP algorithm is first compared with the general purpose constrained
optimization solver $\mathtt{fmincon}$ \cite{ColemanBranchGrace1999}
in MATLAB. From Figure \ref{fig:Convergence-comparison}, it is easy
to see that SCA-MRP obtains a faster convergence and converges to
a better solution than $\mathtt{fmincon}$. We further compare the
SCA-MRP with the inner problem solved by $\mathtt{CVX}$, $\mathtt{MOSEK}$,
and ADMM. The inner problem solved by ADMM can uniformly get a faster
convergence either using acceleration or not than the others. In Figure
\ref{fig:MR-Var}, we show that by tuning the parameter $\mu$ in
problem \eqref{eq:problem formulation}, our formulation is able to
get a trade-off between MR and variance of the portfolio. However,
this desirable property cannot be attained with existing methods in
the literature.

\section{Conclusions \label{sec:Conclusions}}

The optimal mean-reverting portfolio design problem arising from statistical
arbitrage has been considered in this paper. We first proposed a general
model for MRP design where a trade-off can be attained between the
MR and variance of an MRP and the investment leverage constraint is
considered. To solve the problem, a SCA-based algorithm is used with
the inner convex subproblem efficiently solved by ADMM. Numerical
results show that our proposed method can generate consistent profits
and outperform the benchmark methods. 

 \bibliographystyle{IEEEtran}
\bibliography{/Users/ziping/Dropbox/Research/1-Report/Reference/RefAll}

\end{document}